\documentclass{elsart}               % The use of LaTeX2e is preferred.

\usepackage{graphicx}          % Include this line if your 
\begin{document}

\def\ni{\noindent}
\def\about{$\sim$}
\def\arcsec{$\,^{\prime\prime}$~}
\def\arcmin{$\,^\prime$~}
\def\deg{$^{\circ}$~}
\def\erg/cm2sec{ergs~cm$^{-2}$~s$^{-1}$}  
\def\ergcm2{ergs~cm$^{-2}$}  
\def\sqdeg{deg$^{-2}$~}  
\def\mdot{$\dot{m}$~}  
\def\X{$\times$~}
\def\Fx{F$_x$~}
\def\Fv{F$_v$~}
\def\FxFv{{F$_x$/{F$_v$}}~}
\def\Lx{L$_x$~}
\def\Lxh{L$_x$(2-10keV)~}
\def\pc3{pc$^{-3}$~}
\def\rc{r$_c$~}
\def\cm-3{cm{$^{-3}$}~} 
\def\km/s{km~s$^{-1}$~}
\def\Pdot{{$\dot{P}$}~}
\def\P/Pdot{P/2{$\dot{P}$}~}
\def\Edot{{$\dot{E}$}~}
\def\rcore{$r_c$~}
\def\X{$\times$}
\def\vdisp{$\sigma_{v0}$}

\newcommand{\lsim }{{\lower0.8ex\hbox{$\buildrel <\over\sim$}}}
\newcommand{\gsim }{{\lower0.8ex\hbox{$\buildrel >\over\sim$}}}

\newcommand{\Msun}{\ifmmode {M_{\odot}}\else${M_{\odot}}$\fi~}
\newcommand{\Rsun}{\ifmmode {R_{\odot}}\else${R_{\odot}}$\fi~}
\newcommand{\Lsun}{\ifmmode {L_{\odot}}\else${L_{\odot}}$\fi~}
\newcommand{\mv}{\ifmmode {m_{V}}\else${m_{V}}$\fi}
\newcommand{\Mv}{\ifmmode {M_{V}}\else${M_{V}}$\fi}
\newcommand{\lopt}{\ifmmode L_{opt} \else $~L_{opt}$\fi}
\newcommand{\loglopt}{\ifmmode{\rm log}~L_{opt} \else log$~L_{opt}$\fi}
\newcommand{\lx}{\ifmmode L_x \else $~L_x$\fi}
\newcommand{\loglx}{\ifmmode{\rm log}~L_x \else log$~L_x$\fi}
\newcommand{\cmsq}{\ifmmode{\rm ~cm^{-2}~} \else cm$^{-2}$\fi~}
\newcommand{\nh}{\ifmmode{\rm N_{H}} \else N$_{H}$\fi}
\newcommand{\fcgs}{\ifmmode {\rm erg~cm}^{-2}~{\rm s}^{-1}\else
erg~cm$^{-2}$~s$^{-1}$\fi} 
\newcommand{\lcgs}{\ifmmode erg~~s^{-1}\else erg~s$^{-1}$\fi}

\begin{frontmatter}
\runtitle{EXIST all sky hard X-ray imaging}  % Running title for regular 

\title{EXIST: All-Sky Hard X-ray Imaging and Spectral-Temporal 
Survey for Black Holes\thanksref{footnoteinfo}} 
% Title, preferably not more  than 10 words.

\thanks[footnoteinfo]{Paper presented at Wide Field Surveys  
meeting. Corresponding author J.~E.~Grindlay. Tel. 617-495-7204. 
Fax:617-495-7356.}

\author[Harvard]{Jonathan E. Grindlay}\ead{josh@cfa.harvard.edu} ~~and 
the EXIST Team    

\address[Harvard]{Harvard-Smithsonian CfA, 
60 Garden St., Cambridge, MA 02138}       % Please supply                           

%\begin{keyword}                           % Five to ten keywords,  
%black holes; active galactic nuclei; coded aperture imaging; 
%solid-state detectors.                    % chosen from the IFAC 
%\end{keyword}                             % keyword list or with the 
                                          % help of the Automatica 
                                          % keyword wizard

\begin{abstract}                          % Abstract of not more than 200 words.
The Energetic X-ray Imaging Survey Telescope (EXIST) is under study 
for the proposed Black Hole Finder Probe, one of the three {\it Einstein
Probe} missions in NASA's proposed Beyond Einstein Program. EXIST would 
have unique capabilities: it would survey the full sky at 5-600 keV 
each 95min orbit with 0.9-5 arcmin, 10$\mu$sec - 45min 
and \about0.5-5 keV resolution 
to locate sources to 10\arcsec and enable black holes 
to be surveyed and studied on all scales. 
With 1y/5$\sigma$ survey sensitivity 
\Fx(40-80 keV) \about5 \X 10$^{-13}$ \fcgs, or comparable to the  ROSAT soft 
X-ray (0.3-2.5 keV) sky survey, a large sample (\gsim2-4 \X 10$^4$) 
of obscured AGN will be identified and a complete sample of accreting 
stellar mass BHs in the Galaxy will be found. The all-sky/all-time 
coverage will allow rare events to be measured, such as possible 
stellar disruption flares from dormant AGN out to \about100 Mpc.  
A large sample (\about2-3/day) of GRBs will be located 
(\lsim10\arcsec) at sensitivities and bandwidths much 
greater than previously and likely yield 
the highest redshift events and constraints on Pop III BHs. 
An outline of the mission design from the 
ongoing concept study is presented. 
\end{abstract}

\end{frontmatter}

\section{Introduction}
Of the wide-field surveys considered at this meeting, 
one is likely widest of all: temporally-resolved 
imaging in hard X-rays (\about5-600 keV)  
of the full sky every 95min. This is possible with a satellite-borne 
coded aperture imaging telescope array with very wide ``fully-coded'' 
field of view that scans with a photon-counting detector array over the 
full sky each orbit. Such is the concept for the Energetic X-ray 
Imaging Survey Telescope (EXIST), originally recommended by 
the Decadal Survey as a possible ISS mission and then 
considered as a Free Flyer (Grindlay et al 2003), 
and now under study to be the Black 
Hole Finder Probe in NASA's {\it Beyond Einstein} Program. The unique 
all-sky/all-time imaging capability of EXIST opens the temporal survey 
window and is highly complementary to LSST (or any Wide-Field OIR 
surveys) and  LISA. The hard X-ray 
band is particularly well matched for detection and study of 
accreting black holes, from stellar mass (\about10\Msun) in 
X-ray binaries to supermassive (\about10$^{7-9}$ \Msun in  
galactic nuclei, which characteristically emit their peak luminosity 
in a broad band around \about100 keV. At energies above \about7 
keV, photoelectric absorption by column densities 
\nh \gsim10$^{23-24}$ \cmsq in galactic nuclei or X-ray binaries 
in the Galaxy becomes negligible and black holes can be seen 
until \nh \gsim 10$^{25}$ \cmsq and they are obscured by Compton 
scattering. 

\vspace*{-0.5cm}

\section{Black Hole Finder Probe Science to EXIST}
The Black Hole Finder Probe (BHFP) should enable the most 
comprehensive survey for black holes, from super massive to stellar, 
yet undertaken. The EXIST concept for BHFP would emphasize three 
primary science objectives, described below along with related 
science goals.  

{\it Revealing obscured or dormant super massive BHs:} 
The all-sky hard X-ray imaging survey would reveal Type 2 (obscured) 
AGN and nearby (\lsim300 Mpc) stellar tidal disruption events in 
normal galaxies. The Type 2 survey will probe the bulk of the 
cosmic X-ray background (CXB), which is only 
\about50-70\% resolved by Chandra and 
XMM at energies 6-8 keV according to Worsley et al (2004), 
who conclude the missing CXB requires a large population of AGN with 
\nh \about10$^{23-24}$ \cmsq and unabsorbed 
\Lx \about10$^{43-44}$ \lcgs primarily at z \about0.5 - 1.5. 
EXIST would provide a full sky sample of these objects 
out to z \about0.3, and more luminous Type 2 Seyferts and QSOs 
with \Lx \about10$^{45}$ and \about10$^{46}$\lcgs out to z \about1 and
3, respectively, to constrain the growth and evolution of 
super massive BHs in the universe. EXIST would provide the first 
full-sky survey for dormant BHs in galactic nuclei from the expected 
disruption of stars. If just 1\% of this accretion 
luminosity is released in a hard X-ray component (as found for 
virtually all Eddington-rate events), some 10-30 events/y should be 
seen out to \about100 Mpc (Grindlay 2004). The significantly 
increased sensitivity  of EXIST vs. the BAT on Swift, both at 
\lsim15keV and \gsim150 keV,  
should enable detection of the prompt \about5 keV burst 
predicted by Kobayashi et al (2004), providing a LISA trigger for 
nearby (\about10Mpc) disruption events.

{\it Measuring the birth of the first BHs:} 
EXIST would conduct the highest 
sensitivity and bandwidth survey for GRBs out to limiting z \about15 
for most burst fluences detected at z \lsim5 by BATSE (Band 2003). The broad 
spectral coverage and very large FoV of EXIST would maximize 
the detection of the extremes of GRBs: those at the highest z 
or low luminosity events and X-ray flashes at lower z. EXIST will
allow the determination of GRB spectra and the peak energy flux,
E$_{peak}$, in a $\nu F_{\nu}$ broad band spectrum, which may allow 
GRBs to be calibrated as standard candles (Ghirlanda et al 2004) and 
thus allow cosmological redshifts directly from GRB spectra. The 
active anti-coincidence side-shields (over 8m$^2$ of CsI) also would
allow the upper energy range for GRBs detected with EXIST to be 
extended to \about10MeV, or significantly beyond the \about600 keV 
limit of the primary imaging detectors (see below), thus enabling 
GRBs from Pop III BHs to be compared directly with those at lower 
redshifts detected at \about10-1000 MeV by GLAST.

\vspace*{-0.3cm}  

{\it Studying BHs in the Galaxy and AGN as Probes:} 
By continually measuring spectra and time variability of accreting BHs 
in brighter AGN and galactic X-ray binaries, EXIST will constrain 
their physics and evolution.  The population of galactic BHs 
accreting from both high and low mass companion stars will be 
measured from both persistent and transient sources. EXIST 
would study the full population of obscured accreting BHs 
suggested by the  INTEGRAL survey of the galactic bulge 
(Revnivtsev et al 2004). BH populations would be 
probed in high mass binaries  
such as the intrinsically absorbed possible sgBe system 
IGR J16318-4848 discovered with INTEGRAL and 
identified (Filliatre and Chaty 2004) with an IR counterpart that 
resembles the peculiar transient  CI Cam. BHs in low mass binaries  
would be seen as X-ray ``novae'', with characteristic hard spectra 
and \Lx(20-100 keV) \about10$^{38}$ \lcgs  and $\tau$ \about10d 
decay time (cf. McClintock and Remillard 2005; MR05), 
could be studied in detail in the Galaxy and LMC/SMC 
and detected as "new" sources out to the Andromeda 
galaxy (super-flares from soft gamma-ray repeaters,    
magnetars, are detected   
out to Virgo). For brighter stellar mass BHs and transients and AGN, 
the continuous and long-duration 
monitoring will allow measures of QPOs that 
appear to correlate with BH mass (MR05) and 
(for AGN vs. Cyg X-1) accretion luminosity (Leighly 2004). For 
non-thermal and jet-dominated BHs such as microquasars and Blazars, 
EXIST spectra and variability coordinated with IR-TeV observations 
will constrain the BH-jet interface. Notably, for Blazars, 
simultaneous EXIST spectra and GeV-TeV spectra from 
GLAST and VERITAS will anchor the underlying  
SSC model (e.g., Katarzynski et al 2004) to enable 
measurement of the cosmic diffuse IR background flux from the observed 
vs. predicted cutoff in the GeV-TeV spectrum. 

\vspace*{-1cm}

\section{Telescope and Detector Concept}
The EXIST Concept Study mission design (Fig. 1) 
would include two large area and FoV coded aperture telescope 
arrays: a High Energy Telescope (HET; 10-600 keV) and a Low Energy 
Telescope (LET; 5-30 keV). 
The HET is an array of 6 x 3 coded aperture telescopes covering a 
a 131\deg \X 65\deg fully-coded/flat-response FoV (and 
153\deg \X 87\deg FWHM FoV) with total detector area of 5.6m$^2$ of 
imaging (1.25mm pixels) CZT. The LET
is 4 arrays of 7 \X 1 coded aperture telescopes covering a 
similar FoV with total detector area of 1.1m$^2$ of imaging 
(0.158mm strips or pixels) Si. The angular resolution of the 
HET and LET are 5\arcmin and 0.9\arcmin, respectively, or well 
below source confusion limits for the AGN surface density 
at \Fx(20-40keV) \gsim5 \X 10$^{-13}$ cgs as well 
as faint hard X-ray sources in the Galaxy, and sufficient to allow 
\gsim5$\sigma$ threshold detections to be centroided to
\lsim10\arcsec to enable correct optical/IR identifications.

\begin{figure}
\begin{center}
\includegraphics[height=6.cm]{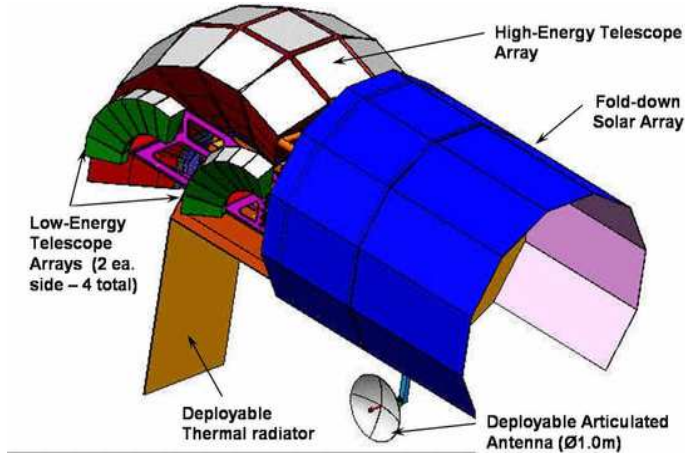}    
\caption{EXIST on orbit. The 131\deg \X 65\deg FoV of the HET  
array is zenith-oriented with the long direction 
perpendicular to the orbit velocity vector and ``nodded'' by 
$\pm$20\deg every \about20min so that the full sky is scanned each
orbit.} 

\label{fig1}                                 % Size the figures 
\end{center}                                 % accordingly.
\end{figure}

%\vspace*{-1cm}
%\begin{ack}                              
I thank the EXIST Team for their role in defining the mission science and 
concept (see http://exist.gsfc.nasa.gov/). 
This work was partly supported by NASA grant NNG04GK33G for 
the EXIST Concept Study.  
%\end{ack}

%\vspace*{-1cm}

\end{document}